\documentclass[usenatbib]{mn2e}
\usepackage{graphicx}
\usepackage{natbib}
\usepackage{amsmath}

\title[2lpt initial conditions] {Second-order Lagrangian perturbation
theory initial conditions for resimulations.}

\begin{document}

\author[A. Jenkins]{Adrian Jenkins\thanks{A.R.Jenkins@durham.ac.uk} \\
Institute for Computational Cosmology, Department of Physics, University of Durham, 
South Road, Durham, DH1 3LE, UK}
\maketitle

\begin{abstract}
   We describe and test a new method for creating initial conditions
for cosmological N-body dark matter simulations based on second-order
Lagrangian perturbation theory (2lpt). The method can be applied to
multi-mass particle distributions making it suitable for creating
resimulation, or `zoom' initial conditions. By testing against an
analytic solution we show that the method works well for a spherically
symmetric perturbation with radial features ranging over more than
three orders of magnitude in linear scale and eleven orders of
magnitude in particle mass. We apply the method and repeat
resimulations of the rapid formation of a high mass halo at redshift
$\sim50$ and the formation of a Milky-Way mass dark matter halo at
redshift zero.  In both cases we find that many properties of the
final halo show a much smaller sensitivity to the start redshift with
the 2lpt initial conditions, than simulations started from Zel'dovich
initial conditions.  For spherical overdense regions structure
formation is erroneously delayed in simulations starting from
Zel'dovich initial conditions, and we demonstrate for the case of the
formation the high redshift halo that this delay can be accounted for
using the spherical collapse model.  In addition to being more
accurate, 2lpt initial conditions allow simulations to start later,
saving computer time.

\end{abstract}


\section{Introduction}

   Computer simulations of cosmological structure formation have been
crucial to understanding structure formation particularly in the
non-linear regime. Early simulations e.g. \cite{Aarseth79} used only
around a thousand particles to model a large region of the Universe,
while recent simulations of structure formation have modelled over a
billion particles in just a single virialised object
\citep{Springel08,Stadel09}.  Since the advent of the Cold Dark Matter
(CDM) model \citep{Peebles82,DEFW85}, most computational effort has been
expended modelling structure formation in CDM universes. Early work in
the 1980s focused on `cosmological' simulations where a representative
region of the universe is modelled, suitable for studying large-scale structure
\citep{DEFW85}. The starting point for these CDM simulations, the initial
conditions, are Gaussian random fields.  The numerical techniques
needed to create the initial conditions for cosmological simulations
were developed in the 1980s and are described in
\cite{Efstathiou85}.

  As the algorithms for N-body simulations have improved, and the
speed of computers increased exponentially with time, it became
feasible in the 1990s to simulate the formation of single virialised
halos in the CDM model with enough particles to be able to probe their
internal structure \citep[e.g.][]{NFW96, NFW97, Ghigna98, Moore99}. These more
focused simulations required new methods for generating the initial
conditions.  The algorithm of \cite{HoffmanRibak91} for setting up
constrained Gaussian random fields was one method which could be
applied to setting up initial conditions for a rare peak where a halo
would be expected to form.  The essence of this technique is that it
allows selection of a region based on the properties of the linear
density field.  However, the objects we actually observe in the
Universe are the end products of non-linear structure formation and it
is desirable, if we want to understand how the structure we see
formed, to be able to select objects on the basis of their final
properties.

  This requirement led to an alternative method for producing initial conditions
based first on selecting objects from a completed simulation
e.g. \cite{Katz93}, \cite{Navarro94}.  The initial conditions for the
first or parent simulation, were created using the methods outlined by
\cite{Efstathiou85}. The density field is created out of a
superposition of plane waves with random phases.  Having selected an
object at the final (or any intermediate) time from the parent
simulation a fresh set of initial conditions with higher numerical
resolution in the region of interest, which we will call
`resimulation' initial conditions (also called `zoom' initial
conditions) can be made. Particles of different masses are laid down
to approximate a uniform mass distribution, with the smallest mass
particles being concentrated in the region from which the object 
forms. The new initial conditions are made by recreating the
the same plane waves as were present in parent simulation together
with the addition of new shorter wavelength power. 

 An alternative technique for creating resimulation initial conditions
was devised by \cite{Bertschinger01} based on earlier work by
\cite{Salmon96} and \cite{Pen97} where a Gaussian random field with a
particular power spectrum is created starting from a white noise
field.  Recently parallel code versions using this method has been developed
to generate very large initial conditions \citep{Prunet08,Stadel09}.
A feature common to both these techniques, to date, is that the displacements 
and velocities are set using the Zel'dovich approximation \citep{Zeldovich70},
where the displacements scale linearly with the linear growth factor.

 It has long been known that simulations starting from Zel'dovich
initial conditions exhibit transients and that care must be taken
choosing a sufficiently high start redshift so that these transients
can decay to a negligible amplitude \citep[e.g][]{Efstathiou85}. A study of the
behaviour of transients using Lagrangian perturbation theory by
\cite{Scoccimarro98}, showed that for initial conditions based on
second-order Lagrangian perturbation theory, the transients are both
smaller and decay more rapidly than first-order, or Zel'dovich, initial
conditions.  \cite{Scoccimarro98}  gave a practical
method for implementing second-order Lagrangian initial
conditions. The method has been implemented in codes for creating
cosmological initial conditions by \cite{Sirko05} and
\cite{Crocce_etal06}.  These codes are suitable for making initial
conditions for cosmological simulations, targeted at large-scale
structure, but they do not allow the creation of resimulation 
initial conditions, the focus of much current work on structure formation.

  In this paper we describe a new method for creating second-order
Lagrangian initial conditions which can be used to make resimulation
initial conditions.  The paper is organised as follows: in
Section~\ref{SECTMOT} we introduce 2lpt theory initial conditions and
motivate their advantages for studying structure in the non-linear
regime by applying them to the spherical top-hat model; in
Section~\ref{MAKERESIMS} we describe how Zel'dovich resimulation
initial conditions are made, and the new method for creating 2lpt
initial conditions, and test the method against an analytic solution
for a spherically symmetric perturbation; in
Section~\ref{REALAPPLICATIONS} we apply the method and analyse the
formation of a dark matter halo at high and low redshift for varying
starting redshifts for both Zel'dovich and 2lpt initial conditions; in
Section~\ref{SECTSUMMARY} we summarise and discuss the main results;
in the Appendix we evaluate two alternate interpolation schemes which
have been used in the process of making resimulation initial conditions.

\section{The motivation for using Second order Lagrangian perturbation 
theory initial conditions\label{SECTMOT}}.

  In this Section we give a brief summary of Lagrangian perturbation
theory (Subsection~\ref{sub2lpt}), concentrating on the second-order
approximation, and illustrate the difference they make for the
spherical top-hat model in Subsection~\ref{substh}.  The method we
describe could be developed further to create third order Lagrangian
perturbation initial conditions.  However, \cite{Scoccimarro98}
concluded that the gains from going from 2nd to 3rd order were
relatively modest while the complexity of computing the third order
terms is considerable.

\subsection{Summary of second-order Lagrangian perturbation theory\label{sub2lpt}}

 We give only a bare-bones account of second-order
perturbation theory here.   Detailed discussions of Lagrangian perturbation
theory in general can be found in the literature
\citep[e.g][]{Buchert94,Buchert_etal94,Bouchet_etal95,Scoccimarro98}.
We follow the notation used in Appendix D1 of \cite{Scoccimarro98}.

  In Lagrangian perturbation theory the Eulerian final comoving
positions {\bf x} are related to the initial positions ${\bf q}$ via a
displacement field, ${\bf\Psi}({\bf q})$:
\begin{equation} 
 {\bf x} = {\bf q} + {\bf\Psi}.
\end{equation}
 
 It can be shown for second-order Lagrangian perturbation theory
(2lpt) that the displacement field is given, in terms of two
potentials $\phi^{(1)}$ and $\phi^{(2)}$, by:
\begin{equation}  \label{twolpt_pos} 
 {\bf x} = {\bf q} - D_1\nabla_q\phi^{(1)} + D_2\nabla_q\phi^{(2)},
\end{equation}
 where $D_1$ is the linear growth factor, and $D_2$ the second
order growth factor, which is given by $D_2 \approx -3D_1^2/7$. The subscripts
${\bf q}$ refer to partial derivatives with respect to the Lagrangian coordinates
${\bf q}$.  Similarly, the particle comoving velocities ${\bf v}$ are given
to second order by:
\begin{equation}  \label{twolpt_vel}
 {\bf v} = -D_1f_1H\nabla_q\phi^{(1)}+D_2f_2H\nabla_q\phi^{(2)}, 
\end{equation}
 where $H$ is the Hubble constant, the $f_i = {\rm d}\ln D_i/{\rm
d}\ln a$ ($a$ is the expansion factor).  For flat models with a
non-zero cosmological constant, the following relations apply
$f_1\approx\Omega^{5/9}, f_2\approx2\Omega^{6/11}$
\citep{Bouchet_etal95}, where $\Omega$ is the matter density. In
practice, these approximations for $f_1$,$f_2$ and $D_2$ are extremely
accurate when applied to making most actual $\Lambda$CDM initial
conditions, as the start redshift is high enough that $\Omega$ is very
close to unity.

  The potentials $\phi^{(1)}$ and $\phi^{(2)}$ are obtained by
  solving a pair of Poisson equations:
\begin{equation}  \label{phi1poisson} 
  \nabla^2_q\phi^{(1)}({\bf q}) =  \delta^{(1)}({\bf q}),
\end{equation}
where $\delta^{1)}({\bf q})$ is the linear overdensity, and 
\begin{equation} \label{phi2poisson}
  \nabla^2_q\phi^{(2)}({\bf q}) =  \delta^{(2)}({\bf q}).
\end{equation}

We will refer to the term $\delta^{(2)}({\bf q})$, as the
`second-order overdensity'.  In fact, this quantity is related to the
linear overdensity field by the following quadratic expression:
\begin{equation}\label{twolpt_source}
 \delta^{(2)}({\bf q})=\sum_{i>j} 
    \Big( \phi^{(1)}_{,ii}({\bf q})\phi^{(1)}_{,jj}({\bf q})-
    [\phi^{(1)}_{,ij}({\bf q})]^2\Big),
\end{equation}
where we use  
$\phi_{,ij} \equiv \partial^2\phi/\partial q_i\partial q_j$ for short.

 The most common technique used in making cosmological initial
conditions, to date, is the Zel'dovich approximation
\citep{Zeldovich70}.  Zel'dovich initial conditions can equivalently be
called first order Lagrangian initial conditions, and are given by
ignoring the terms with $\phi^{(2)}$ in equations~(\ref{twolpt_pos})
and (\ref{twolpt_vel}).  A variant of Zel'dovich initial conditions was
advocated by \cite{Efstathiou85}. In this case the velocities are not
set proportional to the displacements, but proportional to the
gravitational force evaluated for the displaced particle
distribution. This method for setting the velocities gives weaker
transients, as discussed in \cite{Scoccimarro98}, although the scaling
of the transients with redshift is unaffected. In this paper we use
the term Zel'dovich initial conditions to mean the velocities are
proportional to the displacements. We note, and we will exploit the
fact later, that the individual terms on the right-hand side of
equation~(\ref{twolpt_source}) are the derivatives of the first order
Zel'dovich displacements which are linear in the potential
$\phi^{(1)}$.

\subsection{The Spherical top hat model. \label{substh}}

  The spherical top-hat model \citep{Peebles80} provides a convenient way to illustrate
the relative merits of Zel'dovich and 2lpt initial conditions.  Apart
from the significant advantage of mathematical convenience, there are
other reasons for considering this special case:
\begin{description}
\item(i) The first collapsed structures in a cosmological simulation
will form from the regions with the highest linear overdensity. In the
limit of high peak height, the regions around the maxima in a smoothed
linear overdensity field become spherical
\citep{BBKS}. As we will see, it is the first structures which are most
sensitive to whether the initial conditions are Zel'dovich or 2lpt.
\item(ii) It is easy to show that the second-order overdensity,
$\delta^{(2)}$, is related to the first order overdensity,
$\delta^{(1)}$, through the inequality:
$\delta^{(2)}\leq\frac{1}{3}(\delta^{(1)})^2$, with equality occurring for
isotropic compression or expansion.  For a fixed linear overdensity,
the second order overdensity is maximised for isotropic compression,
so the spherical top-hat is the limiting case where the affects of
second-order initial conditions are maximised.
\end{description}

  For convenience we will consider, for now, the top-hat in an
Einstein-de Sitter universe.  In this case the spherical top-hat
solution is most elegantly expressed in a parametric form as
follows. For an overdense spherical top-hat which begins expanding at
$t=0$ and reaches a maximum physical radius of expansion, $R_0$, and
collapses back to zero radius at time, $t_c$, the parametric solution
is:
\begin{equation}\label{parametric}
 r =\frac{R_0}{2}\left(1-\cos\eta\right)\; ; \;
 t = \frac{t_c}{2\pi}\left(\eta-\sin\eta\right),
\end{equation}
where $0\leq\eta\leq 2\pi.$
  These equations can be derived by solving the 
equation of motion for the top-hat:
\begin{equation}\label{eom}
 \ddot{r} = -\frac{\pi^2 R_0^3}{2t_c^2}\frac{1}{r^2},
\end{equation}
 and applying the appropriate boundary conditions.

 If we were to solve the spherical top-hat problem instead by the
techniques used in cosmological numerical simulations, we would begin
by creating initial conditions consisting of a set of particles with
given masses, positions and velocities, at a time, $t$, soon after the
start of the expansion.  The solution would be obtained by evaluating
the gravitational interactions between the particles and integrating
the equations of motion of for all the particles forward in time.  For
a real simulation the accuracy of the subsequent solution would depend
not only on the accuracy of the initial conditions, but also on
numerical aspects such as the time-integration scheme and the force
accuracy. For our purposes, we will assume that these numerical sources
of error can be overcome, and will only consider the effects of using
either Zel'dovich or 2lpt initial conditions on the subsequent
evolution of the top-hat.

  The Zel'dovich and 2lpt initial conditions needed to start a cosmological
simulation of a top-hat are most easily obtained using the rather less
elegant power-series representation of the spherical top-hat solution.
The power-series in $t$ corresponding to equation~(\ref{parametric}) is:
\begin{equation}\label{roft}
 r = \frac{R_0}{4}\left[ \left(\frac{12\pi t}{t_c}\right)^\frac{2}{3}\!\!\!\! -
\frac{1}{20}\left(\frac{12\pi t}{t_c}\right)^\frac{4}{3}  
\!\!\!\!-\frac{3}{2800}\left(\frac{12\pi t}{t_c}\right)^{2}  \ldots\right],
\end{equation}
and the corresponding velocity, $\dot{r}$ is:
\begin{equation}\label{voft}
 \dot{r} = \frac{2\pi R_0}{t_c}\!\!\left[ \left(\frac{12\pi t}{t_c}\right)^{-\frac{1}{3}}
\!\!\!\!\!\!-\frac{1}{10}\left(\frac{12\pi t}{t_c}
\right)^\frac{1}{3}\!\!\!\!\!\!-\frac{9}{2800}\left(\frac{12\pi t}{t_c}\right) \ldots\right].
\end{equation}

  The leading order term in equation~(\ref{roft}) represents the expansion of a
perturbation with mean cosmic density which grows in time in the same
way as the scale factor for an Einstein-de Sitter universe.  The second to
leading 
term gives the linear growth of an overdense spherical growing mode perturbation
and it is easy to verify from equation~(\ref{roft}) that the linear
overdensity at $t=t_c$, when the non-linear solution collapses to a
point, is given by $\delta_c = 3(12\pi)^{2/3}/20) \simeq 1.686$,
familiar from Press-Schechter theory \citep{PS74}.

 We obtain the Zel'dovich initial condition for the spherical top-hat
by taking just the first two leading terms in $r$ and $\dot{r}$ in
equations~(\ref{roft}) and (\ref{voft}).  The equivalent 2lpt initial
conditions correspond to taking the first three terms instead.  We can
follow the top-hat solution starting from these initial conditions, up
until the point when the sphere collapses to zero radius, by
integrating the equation of motion equation~(\ref{eom}) forward in time
from time, $t$, using the values of $r$ and $\dot{r}$ at time $t$ as
the boundary conditions.

\begin{figure}
\resizebox{\hsize}{!}{
\includegraphics{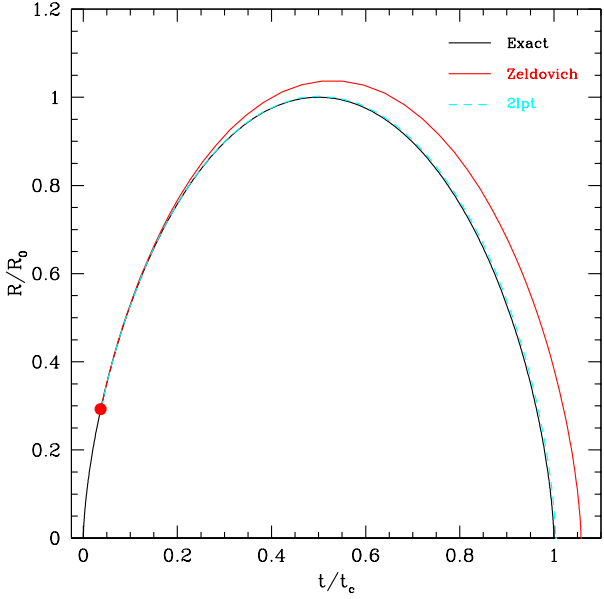}}
\caption{Evolution of a top-hat spherical perturbation. The black
curve shows the radius of a spherical top-hat perturbation as a
function of time for an Einstein-de Sitter universe.  The analytic
solution starts expanding at $t=0$ and collapses at a time $t = t_c$
and has a maximum physical radius of expansion of $R_0$.  The red and
dashed cyan curves show the spherical top-hat solution obtained by
integrating the equation of motion for the top-hat starting from $t =
t_c/27$ and respectively initialising the radius and its velocity with
the Zel'dovich and 2lpt initial condition. The result for the 2lpt
initial condition is more accurate than for the Zel'dovich initial
condition where the time of collapse is significantly delayed.}
\label{tophatfig}
\end{figure}

  In Figure~\ref{tophatfig} the black solid line shows the parametric
top-hat solution given by equation~(\ref{parametric}).  The red solid
curve shows a spherical top-hat model integrated forward from the
arbitrarily chosen time, $t=t_c/27$, starting from the Zel'dovich
initial condition (marked with a red dot).  The cyan dashed curve
shows the corresponding result starting with the 2lpt initial
condition at $t=t_c/27$.  The solution using the 2lpt initial
condition is very much closer to the analytic solution than the curve
started from the Zel'dovich initial condition.

 The collapse time for the top-hat for both the Zel'dovich and 2lpt
initial conditions occurs later than the analytic solution.  This
delay, or timing error, $\Delta t$, can be quantified as follows.
Multiplying equation~(\ref{eom}) by $\dot{r}$ and integrating gives an
integral of the motion: $E = \dot{r}^2/2-\pi^2R_0^3/2t_c^2r.$ For the
analytic solution this can easily be evaluated at the radius of
maximum expansion which gives $E = -\pi^2R_0^2/2t_c^2$.  The
corresponding values of $E$ for the Zel'dovich and 2lpt solutions can
be obtained by substituting the first two or three terms respectively
from equation~(\ref{roft}) and equation~(\ref{voft}) into the
expression for $E$. The expansion time, $t_c$, of the top-hat over the
complete cycle, by analogy with Keplerian orbits, is related to $E$
by: ${\rm d}\ln t_c/{\rm d}\ln|E|=-\frac{2}{3}$ for bound motion
in an Einstein-de Sitter universe.  We can generalise this result to
flat universes with a cosmological constant by numerically integrating
equation~(\ref{eom}) with an added term for the cosmological
constant. In this case we find, by fitting to the results of numerical
integration, an equivalent approximate relation: ${\rm d}\ln t_c/{\rm
d}\ln|E|=-\frac{2}{3}\Omega_c^{0.23}$, valid to an fractional accuracy
of less than 1.2 percent for $0.25 < \Omega_c \le 1$, where $\Omega_c$
is the matter density in units of the critical density at the time of
collapse, provided $t_s\ll t_c$ so the cosmological constant is
negligible at $t_s$.  

 Knowing ${\rm d}\ln t_c/{\rm d}\ln|E|$ and assuming again the condition that
the starting time of the initial condition, $t_s$, obeys, $t_s\ll
t_c$, allows the fractional timing error, $\Delta t/t_c$, to be
obtained, to leading order in $t_s/t_c$, for both types of initial
condition, from the ratio of $E$ for the analytic solution with
respectively with the value of $E$ for Zel'dovich or 2lpt initial
conditions.

 Applying these results we find after some algebra that the timing
error for Zel'dovich initial conditions to lowest order is:
\begin{equation}\label{zeld_delay}
  \frac{\Delta t}{t_c} = \frac{3\delta_c}{10}\left(\frac{t_s}{t_c}\right)^{2/3}\Omega_c^{-0.23}.
\end{equation}
  Similarly, the 2lpt initial
condition fractional timing error is given by:
\begin{equation}\label{2lpt_delay}
  \frac{\Delta t}{t_c} = \frac{23\delta_c^2}{210}\left(\frac{t_s}{t_c}\right)^{4/3}\Omega_c^{-0.23}.
\end{equation}

 Alternatively, these results can be expressed in terms of the start
redshift of a numerical simulation, $z_s$.  Relative to an analytic
spherical top-hat with collapse occurring at redshift, $z_c$, where 
$z_c \ll z_s$, the 
fractional timing error is given for Zel'dovich initial conditions by:
\begin{equation}\label{zeld_te}  \frac{\Delta t}{t_c} 
\approx \frac{0.51}{\Omega_c^{0.36}}\left(\frac{1+z_c}{1+z_s}\right),
\end{equation}
and for  2lpt initial conditions it is:
\begin{equation}\label{lpt_te}
  \frac{\Delta t}{t_c} 
\approx \frac{0.31}{\Omega_c^{0.49}}\left(\frac{1+z_c}{1+z_s}\right)^2,
\end{equation}
 where we have made use another approximate relation:
$t^{2/3}(1+z)\propto\Omega^{0.13}$ relating redshift, $z$, to the age
of the universe, $t$ and the value of $\Omega$ all at time, $t$, which
is accurate to better than one percent in the range $0.25 < \Omega \le
1$ for a flat universe with a cosmological constant.

  Ignoring the weak $\Omega$ dependence, the timing error for the 2lpt
initial condition shows a quadratic scaling with the ratio of the
expansion factor at the start to that at the time of collapse, while
for Zel'dovich initial conditions, the scaling is just linear.  These
results for spherical collapse are consistent with the more general
behaviour of the decay of transients discussed in
\cite{Scoccimarro98}.  The absolute value of the timing error, $\Delta
t$, for a fixed start redshift, actually decreases with increasing
collapse time for 2lpt initial conditions. For Zel'dovich initial
conditions, the timing error grows slowly with increasing collapse
time.  However, by starting at sufficiently high redshift, the
timing error for the Zel'dovich initial conditions can be made small.
It is common practice to start simulations from Zel'dovich initial
conditions at a high redshift for this reason so that the transients
have sufficient time to decay to a negligible amplitude.

Equating the fractional timing errors for equations
equation~(\ref{zeld_te}) and equation~(\ref{lpt_te}) gives an
objective way of defining an equivalent starting redshift for
Zel'dovich or 2lpt initial conditions for a given choice of collapse
redshift.  To be conservative, it would be natural to choose the
collapse redshift to be the first output redshift of the simulation
which is of scientific interest.  To give a concrete example, ignoring
the $\Omega$ dependence, if the 2lpt initial conditions are created a
factor of 10 in expansion before the first output, the equivalent
Zel'dovich initial conditions would need to start at a factor of about
160 in expansion before the first output. Using 2lpt initial
conditions instead of Zel'dovich initial conditions can therefore lead
to a significant reduction in computer time because the simulations
can be started later. A later start has further advantages as force
errors, errors due to the discreteness in the mass distribution and
time integration errors all accumulate as the ratio of the final
expansion factor to the initial expansion factor increases.  All of
these reasons make 2lpt initial conditions attractive for the
simulator.  In the next section we describe the new method to make
2lpt initial conditions.

\section{Making second order Lagrangian resimulation initial conditions\label{MAKERESIMS}}

  In this Section we explain what resimulation initial conditions are
and describe how they can be made.  We will describe the method
currently in use to make initial conditions for the Virgo Consortium
for Supercomputer Simulations\footnote{http://www.virgo.dur.ac.uk/}.
The Virgo resimulation code to date generates Zel'dovich
initial conditions. In Subsection~\ref{set_up} we describe the process
for making these Zel'dovich resimulation initial conditions. In
Subsection~\ref{generation_2lpt} we describe the new method to generate
2lpt initial conditions.  In Subsection~\ref{tests_2lpt} we demonstrate
by comparing with an analytic solution that the method works in
practice.

\subsection{Setting up Zel'dovich resimulation initial conditions \label{set_up}}

 The current method used by the Virgo code is based on the ideas first
described in \cite{Navarro94}.  The actual implementation has changed
considerably over time with increasingly up-to-date accounts
being given in  \cite{Power03}, \cite{Navarro04} and \cite{Springel08}.

 The need for resimulation initial conditions stems from the desire
to repeat a particular simulation of structure formation with
(usually) higher mass resolution in one or more regions than was
present in the original, or parent, simulation.  The parent simulation
may have been run starting from cosmological initial conditions or it
may be a resimulation itself. The following list gives the steps needed
to create a set of resimulation initial conditions from a parent
simulation.

\begin{description}
\item(i)  Identify the region(s) of interest in the parent simulation
          where higher mass resolution is required.

\item(ii) Create a force-free multi-mass particle distribution which
          adequately samples the high resolution region(s) of interest
          identified in (i) with low mass particles. This particle
          distribution needs to be a good approximation to an
          isotropic and homogeneous mass distribution. The initial
          density fluctuations are created by perturbing the positions
          and assigning velocities to each of the particles.  The
          region exterior to the high resolution region can often be
          sampled more crudely than the parent simulation, using larger
          mass particles, as its only purpose is to ensure that the
          gravitational tidal field in the high resolution region is
          accurately reproduced. Typically the masses in the exterior
          region scale roughly as the cube of the distance from
          the high resolution region. We will refer to this
          special particle distribution from now on as the `particle load'.
          
\item(iii) Recreate the displacement and velocity fields for the
          parent simulation and apply the displacements and assign the
          velocities to the particle load created in (ii). 

\item(iv) Add extra small-scale fluctuations to the region of
          interest only, to account for the fact that the mass
          distribution in the resimulation is now more finely sampled
          than the parent simulation and therefore has a higher
          spatial Nyquist frequency.  For Zel'dovich initial conditions
          this high spatial frequency power can be added linearly to
          the low frequency power copied from the parent 
          simulation.
\end{description}

 In this paper we will not be concerned with steps (i) and (ii), 
which are common to resimulations in general, but will concentrate on
adapting steps (iii) and (iv) to allow the creation of 2lpt initial
conditions.

  The displacement and velocity fields needed to perturb the particle
load can in practice be efficiently
created using Fourier methods.  Adopting periodic boundary conditions
when setting up the perturbations forces the Fourier representation of
the density, displacement and velocity fields to be composed of
discrete modes in Fourier space.  The amplitudes and phases of these
modes are set as described in \cite{Efstathiou85} to create a Gaussian
random field with the appropriate power spectrum in Fourier space.
Starting from the Fourier representation of the density field it is
straightforward to use discrete Fourier transforms to generate each of
the components of the displacement field in real space. The discrete
Fourier transform gives the displacements on a cubic lattice.

  In practice, because computer memory is limited, more than one
Fourier grid is required to be able to complete step (iv).
These grids have different physical sizes, with the largest grid
covering the entire periodic cubic volume, as is the case when making
cosmological initial conditions.  The additional grid(s) are given a
smaller physical size, which means even with limited computer memory,
that they can generate a displacement field with power at higher
wavenumbers than the largest grid, but only over a restricted part of
the simulation volume.  By nesting sufficient Fourier grids about
the high resolution region it is possible to generate power all they
way down to the particle Nyquist frequency in the high resolution
region. A description of the creation of the initial conditions
for the Aquarius halos, which required two Fourier grids, is
given in \cite{Springel08}.

For Zel'dovich initial conditions, the displacement field at any
location is formed by coadding the displacement contributions from the
one or more of these overlapping cubic Fourier grids, and the
corresponding velocity field is simply proportional to the
displacement field.  For each Fourier grid only a subset of the
possible modes are assigned non-zero values.  In k-space these
`populated' modes are split between the different grids in a nested
fashion around the origin. Each grid populates a disjoint region of
k-space concentric on the origin, and taken together the modes from
all the grids fill a spherical region in k-space around the origin.
The smallest physical grid, which typically fits around just the high
resolution region, is used to represent the highest wavenumber modes
in k-space. The highest wavenumber possible is limited by the particle
Nyquist frequency which is determined by the interparticle separation.

A displacement field is generated on each Fourier grid. The complete
displacement field is calculated by adding the the contributions from
the various grids together.  To do this it is necessary to define a
set of common positions at which to co-add the displacements.  The
most natural common positions to use are those of the particle load
itself, as the displacements need to be known at these locations to
generate the Zel'dovich initial conditions. The displacements at the
particle positions are computed by interpolating the values from the
nearest Fourier grid points.  A discussion of the interpolation
methods currently in use and their associated errors is given for
completeness in the Appendix.  The interpolation errors only become
significant for fluctuations with a wavelength comparable to the
Nyquist wavelength of the Fourier grid.

 For the high resolution region a second scale, the particle Nyquist
wavelength also becomes important.  Even in the absence of
interpolation errors, the discreteness of the mass distribution
affects the growth of linear fluctuations with wavelengths close to
the particle Nyquist frequency \citep{Joyce05}. We will not be
concerned about either of these sources of error in the main part of
the paper.

\subsection{Generating 2lpt resimulation initial conditions\label{generation_2lpt}}

 In Subsection~\ref{set_up}, we outlined a method for making
Zel'dovich initial conditions for resimulations.  In this Subsection
we will take this method for granted, and concentrate on how to
augment the Zel'dovich initial conditions to produce 2lpt initial
conditions. This is achieved, as can be seen from equations
~(\ref{twolpt_pos}) and (\ref{twolpt_vel}), by computing the
components of the field $\nabla_q\phi^{(2)}$. The second-order corrections
to the Zel'dovich initial conditions for both the positions
and the velocities are proportional to this quantity.

   The second-order potential, $\nabla_q\phi^{(2)}$, is obtained by
solving the Poisson equation~(\ref{phi2poisson}), where the source
term is the second-order density field, $\delta^{(2)}$, given by
equation~(\ref{twolpt_source}).  The right-hand side of this equation
is a non-linear combination of six distinct terms, each of which are
second derivatives of the first order potential, $\phi^{(1)}$.  This
same first order potential is used to generate the Zel'dovich
displacements and velocities. For this reason, as pointed out by
\cite{Scoccimarro98}, the calculation of the second-order overdensity
field, fits in very well with the generation of Zel'dovich
resimulation initial conditions, from a computational point of view.
The six second derivatives can be calculated using Fourier methods in
the same way as is used to compute the first derivatives of
$\phi^{(1)}$ needed for the Zel'dovich approximation.

 For initial conditions that can be made using a single Fourier grid,
the six derivatives can be combined to give the value of
$\delta^{(2)}$ at each Fourier grid point.  The Poisson 
equation~(\ref{phi2poisson}), can then be solved immediately using
standard Fourier methods \citep{Hockney88} to give the components of
$\nabla_q\phi^{(2)}$ at the grid points. Choosing a suitable cubic
grid arrangement for the particle load ensures that the Lagrangian
positions of the particles coincide exactly with Fourier grid points.
This means the particle 2lpt displacements and velocities can be
evaluated without the need for any spatial interpolation.  This is the
approach adopted by the 2lpt codes made available \cite{Sirko05} and
\cite{Crocce_etal06}.

 The approach just described does not work when more than one Fourier grid
is used to generate the Zel'dovich initial conditions, because
$\delta^{(2)}$ is a non-linear function of the first-order potential,
$\phi^{(1)}$, unlike the Zel'dovich displacements and velocities which
are linear functions.  The solution to this problem is simply to
compute all six of the derivatives, which are linear functions of
$\phi^{(1)}$, for each Fourier grid, and co-add them. Then once the summation is
complete for all six derivatives for a given particle, compute $\delta^{(2)}$
using equation~(\ref{twolpt_source}).

  As with the Zel'dovich displacements, the values of the six
derivatives need to be interpolated to a common set of positions
before they can be co-added. It proves convenient again to use the
particle load to define these common positions.  Computationally this
entails storing six additional quantities per particle in computer
memory.  We use the QI interpolation scheme, described in the
Appendix, to interpolate the derivatives to the positions given by the
particle load, as this scheme helps to minimise the interpolation
errors, given limited computer memory.  We
have implemented the algorithm to compute $\delta^{(2)}$ for each
particle in the particle load in a Fortran/C code called
{IC\_2LPT\_GEN}. This code also computes the Zel'dovich displacements
and velocities for each particle at the same time.

The final stage to make the 2lpt resimulation initial
conditions is to solve the Poisson equation~(\ref{phi2poisson}). At
this point it is no longer trivial to solve the equation by Fourier
methods because the second-order overdensity field is known only at
the positions of the particle load, which do not in general form a
regular grid.  However it is relatively simple to solve the Poisson
equation using an N-body code Poisson solver.  It is natural, given
that the purpose of creating the initial conditions in the first place
is to integrate them forward in time with an N-body code, to use the
same N-body code to solve both equation~(\ref{phi2poisson}) and to
integrate the equations of motion.

 As mentioned in Subsection~\ref{set_up}, in point (ii), the
Lagrangian positions of the particle making up the particle load are
special.  The particles are distributed so that the mass density field
is an excellent approximation to a uniform mass density, on scales
larger than the interparticle separation. Furthermore the
gravitational force on any particle due to the others cancels out to a
very good approximation.  We can exploit the properties of the
particle load in a simple way to reconstruct the second-order density
field.  To do this we create a `2lpt particle load' by taking the
positions from the particle load, but replacing the particle masses
with a `2lpt' mass, given by:
\begin{equation}   
 m_{\rm 2LPT} =  m(1 + \lambda\delta^{(2)}),
\label{2lpt_masses}\end{equation} 
where $\lambda$ is a constant numerical factor, whose value for now we
will take to be unity. 

 With minimal modification, an N-body code can then be used to compute
the gravitational force due to the 2lpt particle load acting on itself.
By construction this amounts to solving equation~(\ref{phi2poisson}),
with the resulting gravitational acceleration giving,
$\nabla_q\phi^{(2)}$, at the positions given by the particle load.
Provided the N-body code also reads in the Zel'dovich velocities for
the particle load, the 2lpt initial condition positions and velocities
can be evaluated using equations~(\ref{twolpt_pos}) and
(\ref{twolpt_vel}) respectively.  Discarding the 2lpt masses, and
replacing them with the true particle masses completes the creation of the 2lpt
resimulation initial conditions.  The N-body code can then be used in
the normal way to integrate the equations of motion for all of the
particles.

   The amplitude of the second-order overdensity field, and similarly
$\nabla_q\phi^{(2)}$ depends on the redshift of the initial
conditions.  By introducing the scaling factor $\lambda$ into the
`2lpt' masses, and rescaling the resulting $\nabla_q\phi^{(2)}$ by
$1/\lambda$ to compensate, it is possible to make the calculation
performed by the N-body code independent of the requested redshift of
the initial conditions.  It is desirable to implement this scaling if
one wants to calculate $\nabla_q\phi^{(2)}$ accurately for high
redshift initial conditions, as otherwise the inevitable force errors
from the N-body force calculation will begin to dominate over the
small contribution from the second-order overdensity field.  The
precise value of $\lambda$ is unimportant, provided it is large
enough.  A suitable choice is given by setting $\lambda$ to be as
large as possible while avoiding any of the `2lpt' masses becoming
non-positive.

 We have implemented the final step to make the 2lpt initial
conditions in practice with the P-Gadget3 code, written by Volker
Springel.  P-Gadget3 is an N-body/SPH code written in C. This code is
similar to the public Gadget-2 code \citep{Springel05b}, and has been
used, for example, for the Aquarius project \citep{Springel08}.
Operationally the Fortran/C code, {IC\_2LPT\_GEN}, on completion
outputs a special format initial condition file(s) for the P-Gadget3
code to read in. The file is special in that the particle positions
are the Lagrangian positions, {\it i.e.} the particle load
positions, the velocities are the Zel'dovich velocities, and an
additional block of data is provided which gives every particle a
second mass - the `2lpt-mass' described above.  This required a
modification to P-Gadget3 so that it first can recognise the special
file format required by the 2lpt initial conditions through a new flag
stored the file header. Having read in the initial conditions,
P-Gadget3 first does a force calculation using the 2lpt masses for
each particle as described above, and once the 2lpt initial conditions
have been created, integrates the equations of motion forward in time
as usual.

  The linear density field created by the resimulation method outlined
above is bandwidth limited (see \cite{Numerical_recipes92} for a
discussion).  The second-order overdensity field is similarly
bandwidth limited but because equation~(\ref{twolpt_source}) is
quadratic, the second-order overdensity has twice as large a bandwidth
in each physical dimension.  If the second-order overdensity field is
known at the Lagrangian particle positions then it is only fully
sampled if the linear density is limited to a wavelength which is
twice that of the particle Nyquist wavelength (this approach is
implemented in the \cite{Crocce_etal06} code).  It is common practice
in making initial conditions to use waves for the linear density field
down to the Nyquist wavelength itself. In this case the second-order
overdensity field at the positions of the particles undersamples the
second-order density field. However, as mentioned earlier, the growth rate of
fluctuations of very short wavelengths is also compromised by the
discreteness of the mass distribution in any case whether using
Zel'dovich or 2lpt initial conditions so we will ignore this issue, but
it remains the case that it is always advisable to do resolution tests
to establish the limitations of the simulations.

\subsection{Testing the 2lpt resimulation initial conditions \label{tests_2lpt}}

 We can test limited aspects of the {IC\_2LPT\_GEN} code against the
code made public by \cite{Crocce_etal06} for the cases of both
Zel'dovich and 2lpt cosmological initial conditions, where the
second-order overdensity field can be computed using a single Fourier
grid.  The two codes were written independently although they do share
some subroutines from \cite{Numerical_recipes92}.  We find that when
the \cite{Crocce_etal06} code is modified so as to set up precisely
the same density field (i.e. exactly the same modes, with the same
phases and amplitudes for each mode, and the same shape linear input
power spectrum, the same power spectrum normalisation and the same
cosmological parameters) then the output of the two codes agrees
extremely well.

     Testing the 2lpt resimulation initial conditions generated using
the method described in the previous section for a general case is
problematic as it is non-trivial to compute the second order terms by
alternative means.  However, it is relatively simple to test the code for
a spherically symmetric perturbation for which one can calculate the
properties of the initial conditions analytically.  The
{IC\_2LPT\_GEN} code needs only minimal alteration to be able to
create a spherical perturbation.  The only change needed is
to replace the part of the code which generates the amplitudes and
random phases for each Fourier mode, as required for generating random
Gaussian fields, with a module that instead computes the appropriate
amplitude and phase for a specified spherical perturbation with a
centre of symmetry at a given location. In other respects the code is
unaltered, which means the spherical test remains a powerful test of the
overall method and the particular numerical implementation.  

The test initial condition most closely resembles what would be
expected around a high overdensity peak and is similar to the initial
conditions in Section~\ref{REALAPPLICATIONS} for the highest
resolution initial conditions used in the papers \cite{Gao05} and
\cite{Reed05}. These two papers were based on a simulation of a relatively
massive and very rare high redshift halo, a suitable location 
for finding one of the first metal free stars.

 For our test we add a spherically symmetric perturbation,
to a homogeneous and isotropic Einstein de-Sitter 
universe. We adopt a simple analytic form for which the
second-order overdensity and the 2lpt displacements and velocities can
be computed analytically.  The perturbation was formed by summing $n$
component terms to give the linear overdensity field as follows:
\begin{equation}\label{spherical_pert}
\delta^{(1)}(r) = \sum_{i=1}^n\delta_i^{(1)}(3-r^2/\sigma_i^2)\exp(-r^2/2\sigma_i^2),
\end{equation}
 where $r$ is the radius, $\delta_i^{(1)}$ is the amplitude of the
linear overdensity, and $\sigma_i$ the linear size, for the $i$th
component of the perturbation. The prefactor to the exponential is
chosen so that: $\int_0^\infty r^2 \delta^{(1)} {\rm d}r = 0,$ meaning
that the net mass of the perturbation is zero.

The choice of a radial Gaussian cut-off ensures that the spherical
perturbation is well localised which is a necessary requirement: the
use of periodic boundary conditions in making the resimulation initial
conditions violates spherical symmetry, and means the choice of the
scale lengths $\sigma_i$ has to be restricted or the resulting
perturbation will not be close to being spherically symmetric.  We
have chosen the scale lengths, $\sigma_i$,  so that each component of the
perturbation is essentially generated by a corresponding Fourier grid,
with the value of $\sigma_i$ being no greater than about 10 percent of
the side length of the grid and each grid is centred on the centre of
spherical symmetry. These requirements ensure that each component of the
perturbation is close to, although not exactly, spherical.

 All that is needed to build the perturbation with the initial conditions code is 
the equivalent Fourier representation of the real space perturbation equation~(\ref{spherical_pert}).
 The Fourier representation, $\hat{\delta}({\bf k})$ for a perturbation
with overdensity $\delta(x)$ is given by: 
\begin{equation}
   \hat{\delta}({\bf k}) = \int\delta(x)\exp(i{\bf k}\cdot{\bf x}){\rm d}^3{\bf x},
\end{equation}
where ${\bf k}$ is the wavevector.
 Equation~(\ref{spherical_pert}) has the following Fourier representation as
a function of the magnitude of the wavevector $k$:
\begin{equation} 
\hat{\delta}(k) = (2\pi)^{3/2}\sum_{i=1}^n\delta_i^{(1)}k^2\sigma_i^5\exp(-k^2\sigma_i^2/2).
\end{equation}
 The solution of equation~(\ref{phi1poisson}) is:
\begin{equation}
\phi^{(1)}(r) =\sum_{i=1}^n\delta_i^{(1)}
\sigma_i^2\exp(-r^2/2\sigma_i^2). 
\end{equation}  
For a general spherical perturbation, the source term for the Poisson
equation~(\ref{phi2poisson}) simplifies in spherical polar coordinates
to:
\begin{equation}
  \nabla^2\phi^{(2)}
     =  \frac{1}{r^2} \frac{{\rm d}}{{\rm d}r} \left( r \left(\frac{{\rm d\phi^{(1)}}}{{\rm d}r}\right)^2\right)
\end{equation}
 which can be integrated to give the second order displacement:
\begin{equation} 
\frac{{\rm d}\phi^{(2)}}{{\rm d}r}=\frac{1}{r}\left(\frac{{\rm d}\phi^{(1)}}{{\rm d}r}\right)^2,
\end{equation}
 where the constant of integration has been set to zero to avoid a source term at
the coordinate origin.

 For the spherical perturbation given by equation~(\ref{spherical_pert}),
the potential
is given by:
\begin{equation}
\phi^{(2)}(r) = \frac{1}{2}\sum_{i=1}^n\sum_{j=1}^n\delta_i^{(1)} \delta_j^{(1)}
\sigma_{ij}^2\exp(-r^2/\sigma_{ij}^2),
\end{equation}
where $2/\sigma_{ij}^2 = 1/\sigma_{i}^2 + 1/\sigma_{j}^2.$

  The second-order overdensity, given by equation~(\ref{twolpt_source}) 
is:
\begin{equation}
 \delta^{(2)}(r) = \sum_{i=1}^n\sum_{j=1}^n\delta_i^{(1)} \delta_j^{(1)}
\left(3-\frac{2r^2}{3\sigma_{ij}^2}\right)\exp(-r^2/\sigma_{ij}^2).
\end{equation}

 For the test, a set of initial conditions was made summing five terms
in Eq.~(\ref{spherical_pert}).  The simulation volume was chosen to be
periodic within a cubic volume of side length 100 units and the
centre of symmetry was placed in the middle of the simulation
volume. Five Fourier grids were used also centred on the centre of
the volume. The parameters chosen are given in
Table~\ref{table_test}. The Table gives the values of $\delta_i^{(1)}$ and
$\sigma_i$ for the five components making up the spherical
perturbation. For the purposes of the test the actual value of the
overdensity is unimportant. We arbitrarily chose the central overdensity to be unity. 
This is an order of magnitude higher than what
would typically occur in a realistic simulation.  The quantity $L_{\rm
grid}$ gives the physical dimension of each Fourier grid. The first grid, which
covers the entire cubic periodic domain, has a dimension of 100 units
on a side. The parameter $N_{\rm grid}$ is the size of the Fourier
transform used in the computations. The parameters $K_{\rm min}$ and
$K_{\rm max}$ define which Fourier modes are used to make the
spherical perturbation.  All modes are set to zero except for the
modes in each grid with wavenumber components, $2\pi/L_{\rm
grid}(k_x,k_y,k_z)$, which obey $K_{\rm min}\le {\rm
max}(|k_x|,|k_y|,|k_z|)<K_{\rm max}$. These limits apply to grids 1, 2, 3, 4
and grid 5, except for the high-k limit, which is replaced by
the condition $\sqrt{k_x^2 + k_y^2 + k_z^2}<K_{\rm max}$ to ensure a
spherical cut-off in k-space to minimise small-scale anisotropy.

Following stage (ii) of the process described in
Subsection~\ref{set_up}, a special particle load was created
for the test consisting of concentric cubic shells of particles
centred on the centre of the volume.  In detail, the cubic volume of
side length 100 was divided into $n^3$ cubic cells. A particle was
placed at the centre of just those cells on the surface of the cube (a
total of $n^3-(n-2)^3$ cells). The mass of the particle was given by
the cell volume times the critical density. The remaining $(n-2)^3$
cells define a smaller nested cube of side length $100\times(n-2)/n$.
This smaller cubic volume was again divided into $n^3$ cells and the
process of particle creation repeated. The process was repeated 360
times, taking $n=51$. On the final iteration all $n^3$ cells were
populated, filling the hole in the middle. Provided the spherical
distribution is centred on the centre of the cube this arrangement of
particles allows a spherical perturbation covering a wide range of
scales to well sampled everywhere.  The particle load  has
cubic rather than spherical symmetry, so this arrangement will
generate some asphericity, but for sufficiently large $n$, this
becomes small. Similar results were obtained with a smaller value,
$n=23$.

\begin{figure}
\resizebox{\hsize}{!}{
\includegraphics{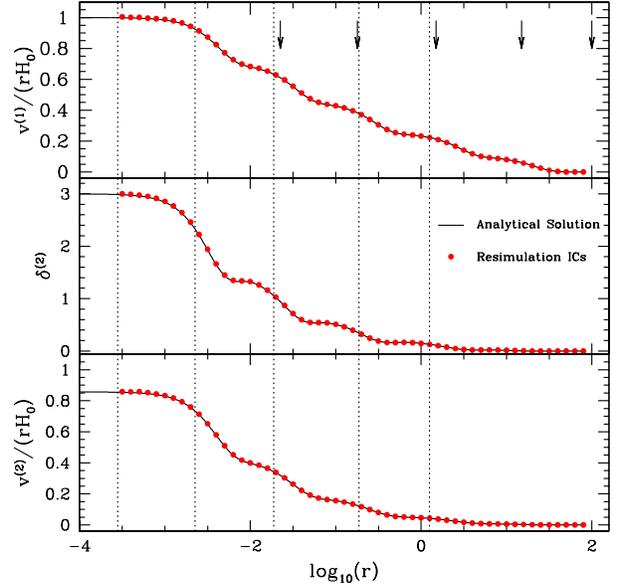}}
\caption{ Comparison between quantities generated by the resimulation
codes for a spherical perturbation and the exact analytical
result. The top panel shows the Zel'dovich velocities ($v^{(1)}$) are
accurately reproduced by the {IC\_2LPT\_GEN} code.  The middle panel
shows the second-order overdensity ($\delta^{(2)}$), output by the
{IC\_2LPT\_GEN} code. This quantity is needed as an input to the {P-Gadget3}
code to generate the second order Lagrangian displacements and
velocities. The bottom panel shows the second order Lagrangian
velocities computed by the P-Gadget3 code using the values of
the second-order overdensity computed for each individual particle.
The arrows mark the side lengths of the five periodic Fourier grids
that are used to generate the perturbation in the resimulation
code. The dotted vertical lines mark the short wavelength cut-off for each of the
Fourier grids.}
\label{tlpt_test_fig}
\end{figure}

 Figure~\ref{tlpt_test_fig} shows the results of the test.  The top
panel shows the Zel'dovich velocities ($v^{(1)}$) as a function of
radius, scaled by the Hubble velocity relative to the centre.  The
black line shows the analytic solution, while the red dots show 
binned values measured from the output of the {IC\_2LPT\_GEN} code.
The vertical arrows mark the edge length of the five cubic Fourier
grids, while the five vertical dotted lines show the high spatial
frequency cut-off (given by $K_{\rm max}$) of the imposed perturbations
for the corresponding cubic grid.  The first panel establishes that
the {IC\_2LPT\_GEN} code is able to generate the desired linear
perturbation.  

 The second panel of Figure~\ref{tlpt_test_fig} shows in black the
second-order overdensity of the analytic solution, while the red dots
show the results from the output of the {IC\_2LPT\_GEN}. This shows
that the second-order overdensity is successfully reproduced by
the code.

The third panel of Figure~\ref{tlpt_test_fig} shows the second-order
velocity ($v^{(2)}$). The black curve is the analytic solution and the
red dots show the binned second-order contribution generated by the
P-Gadget3 code, using the values of the second-order overdensity
computed for each particle which are generated by the {IC\_2LPT\_GEN}
code.  The method works well, the fractional error in the value of
$v^{(2)}$ is below 1 percent everywhere except close to the scale of
the first grid where spherical symmetry is necessarily broken and the
perturbation has an extremely small amplitude anyway.

  In conclusion the method is able to work well for a perturbation
that spans a very large range of spatial scales and particle masses.
The 5th component has a length-scale of just 1/5000th of the 1st
component and the particle masses associated with representing these
components vary by the cube of this number - more than eleven
orders of magnitude.  Most practical resimulations are less demanding,
although the resimulation of \cite{Gao05}, which we reproduce in
the next Section, is similar.

\begin{table}
  \begin{center}
{
\begin{tabular}{|l|l|l|l|l|l|l|} 
 \hline
 $i$      &  $\delta_i^{(1)}$  &  $\sigma_i$  &  L$_{\rm grid}$ & N$_{\rm grid}$ & K$_{\rm min}$ &  K$_{\rm max}$  \\
\hline
  1       &      0.1           &     15     &      100       &    512            & 1             &  80              \\
  2       &      0.15          &      2     &       15       &    512            & 12            &  80              \\
  3       &      0.2           &      0.2   &       1.5      &    512            & 8           &  80              \\
  4       &      0.25          &      0.024  &       0.18      &   512            & 10           &  80              \\
  5       &      0.3           &      0.003 &     0.0225    &    512            & 10           &  80              \\
\hline
\end{tabular}
}
\end{center}
\caption{Parameters for a spherical perturbation used to test the
 method for generating 2lpt initial conditions. The index $i$ and
 quantities $\delta_i^{(1)}$ and $\sigma_i$ characterise the
 perturbation used for the test as given by
 equation~(\ref{spherical_pert}).  The remaining columns give
 parameters of the Fourier grids used to realise the perturbation and
 are explained in the main text.}
\label{table_test}
\end{table}

\section{Real applications\label{REALAPPLICATIONS}}

 In this Section we remake some resimulation initial conditions with
Zel'dovich and 2lpt initial conditions and run them with the P-Gadget3
code. We investigate how the end point of the simulation depends on 
the start redshift and the type of initial condition.

\subsection{Structure formation at high redshift}

  The initial conditions for the spherically symmetric test in
Subsection~\ref{tests_2lpt}, was intended to approximate the initial
conditions in a series of resimulations presented in \cite{Gao05} to
investigate the formation of the first structure in the Universe.  In
that paper, a cluster was selected from a simulation of a cosmological
volume.  A resimulation of this cluster was followed to $z=5$, and the
largest halo in the high resolution region containing the
proto-cluster was itself resimulated and followed to $z=12$. The
largest halo in the high resolution region at this redshift was then
resimulated and followed to $z=29$. This was repeated one more time
ending with a final resimuation (called `R5') yielding a halo of mass
$M_{200} = 1.2\times10^5h^{-1}M_\odot$ (that is the mass within a
spherical region with a mean density of 200 times the critical
density) at redshift 49, simulated with particles with a mass of just
$0.545h^{-1}M_\odot$. The original motivation for studying such a halo
is that halos like this one are a potential site for the formation of
the first generation of metal free stars \citep{Reed05}.

\begin{figure}
\resizebox{\hsize}{!}{
\includegraphics{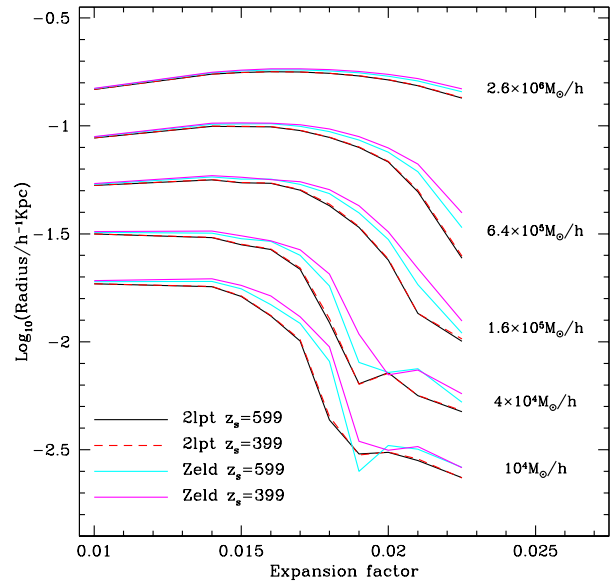}}
\caption{The physical radius of a sphere containing a fixed amount of mass 
in a simulation of the formation of a massive halo at high redshift.
 The sphere is centred on a density maximum located by applying the
shrinking sphere algorithm detailed in the text. The four curves show
simulations starting from either Zel'dovich or 2lpt initial conditions,
at a start redshift of 399 or 599.  The agreement between the two 2lpt
initial conditions is very good. As expected from the spherical
collapse model, discussed in Subsection~\ref{substh}, the collapse of the 
Zel'dovich initial conditions is delayed with the largest delay occurring
for $z_s=399.$}
\label{figleft}
\end{figure}

\begin{figure}
\resizebox{\hsize}{!}{
\includegraphics{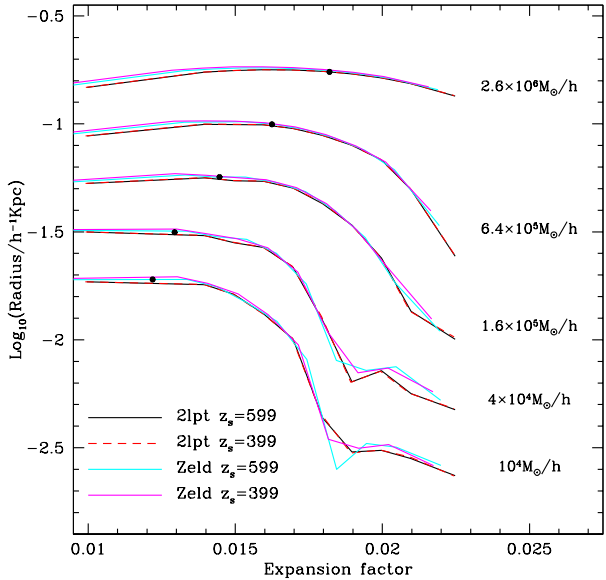}}
\caption{ Same as Figure~\ref{figleft} except that the expansion
factor plotted for each of the four curves has been corrected to
account for the time delay expected in the spherical collapse model
and given by equations~(\ref{zeld_te}) and (\ref{lpt_te}). The
density inside each radius exceeds ten times the critical density to
the right of the black dots. The correction is expected to work well
for a spherical collapse above an overdensity of around ten (see main
text).  Even through the collapse is not perfectly spherical
applying the correction does greatly improve the match between the four curves
particularly to the right of the black dots. The spacing of the outputs is insufficient
to resolve features around $a=0.018$ where shell crossing becomes important
in the lower two curves. }
\label{figright}
\end{figure}

   We have recreated the R5 initial conditions with both Zel'dovich and
2lpt resimulation initial conditions.  In \cite{Gao05}, the starting
redshift was 599, and we have repeated a simulation starting from this
redshift and also $z_s=399$, for both Zel'dovich and 2lpt initial
conditions.  We ran each simulation with P-Gadget3 to $z=43\frac{1}{3}$ and
produced outputs at expansion factors of 0.01, 0.015, 0.017, 0.019,
0.021, 0.0225.  These initial conditions remain the most complex that have been
created by the Virgo resimulation code, requiring seven Fourier
grids. The very large dynamic range also required that the simulation
of R5 itself was carried out with isolated boundary conditions
following a spherical region of radius $1.25h^{-1}$Mpc which is much
smaller than the parent simulation of side length $479h^{-1}$Mpc.  We
adopted the same isolated region for the new simulations. 

The use of isolated boundary conditions gives the isolated region as a
whole a velocity which depends both on the starting redshift and
initial condition type.  In order to compare the different versions of
R5 at particular redshifts it is first necessary to establish a
reference point or centre from which to measure.  Using a position
determined from the main halo is one potential approach, but it was
found in \cite{Gao05} that using a group finder like
friends-of-friends on this rather extreme region results in groups
that are highly irregular in shape, and are therefore unsuitable.
Instead we have found a centre for each simulation output using the
shrinking sphere algorithm described in \cite{Power03} starting with a
sphere that encloses the entire isolated region.  Having found a
centre, we have grown spheres from this point to find the radii
enclosing masses of $(10^4, 4\times10^4, 16\times10^4, 64\times10^4,
256\times10^4)h^{-1}M_\odot$. We checked, for one output redshift, by
making dot plots that this centre can be identified visually as being
in the same recognisable lump.
 
  Figure~\ref{figleft} shows these five radii as a function of
the expansion for the four resimulations.  The radii for the two
resimulations starting from 2lpt initial conditions at redshifts 399
and 599 are extremely close. The simulations starting from Zel'dovich
initial conditions show a significantly delayed collapse, with the
lower start redshift showing the largest delay.  This is in accordance
with what would be expected from the predictions of the spherical
collapse model as shown in Subsection~\ref{substh}.

The collapse occurring in the R5 simulation is only approximately
spherical as evidenced by Figures 3 and 4 in \cite{Gao05}. It is
nonetheless interesting to see whether the results from the spherical
top-hat model do agree quantitatively as well as qualitatively. To do
this we first estimated an expansion factor of collapse for each of
the five 2lpt $z_s=599$ curves, by fitting a spherical top-hat to each
curve, and taking a value for the expansion factor of collapse from
the corresponding top-hat model. We chose these particular curves
because the expected timing error for these 2lpt initial conditions
should be very small, compared to the simulations started from
Zel'dovich initial conditions, and so should not bias the estimate of
the collapse expansion factor significantly.  In Figure~\ref{figright}
the same radii are plotted as in Figure~\ref{figleft}. However the
values of the expansion factors along the x-axis are `corrected' using
the equations~(\ref{zeld_te}) and (\ref{lpt_te}) to cancel out the
expected timing errors for both the 2lpt and Zel'dovich initial
conditions. The timing error is assumed constant along each curve.
Strictly speaking this correction applies at the moment the spherical
top-hat solution collapses to a point. However inspection of
Figure~\ref{tophatfig}, together with the fact that top-hat solution
is symmetric in time, for any initial condition, about the point of
maximum expansion (for an Einstein-de Sitter universe, an excellent
approximation at redshift $\sim50.$), shows that the timing error
mostly arises around the time of maximum expansion and therefore the
correction at the time of collapse, should be applicable once the
top-hat has collapsed by more than $\sim0.85$ of the radius of maximum
expansion.  For the spherical top-hat this corresponds to the density
exceeding the critical density by a factor of about ten, and this
overdensity is marked on each curve in Figure~\ref{figright} by a
black dot. To the left of the dots the density is less than ten so the
correction is expected to be too strong, which it is.  To the right of
the dots the shift works rather well with the lines showing much
better agreement than in Figure~\ref{figleft}.  The spacing of the
outputs is too crude to follow the collapse of the innermost two radii
very accurately, so it is not surprising the agreement around
$a=0.018$ is less good than elsewhere.

  In conclusion, the results using 2lpt initial conditions are much
less sensitive to the start redshift and it is possible to start the
simulation later. The formation of the halo is delayed using Zel'dovich
initial conditions and the magnitude the delay agrees with what one
would expect for a spherical top-hat collapse.

\subsection{Aquarius halos from 2lpt initial conditions}

 As a final example, we look at the effect of using 2lpt resimulation
initial conditions on the redshift zero internal structure of one of
the Aquarius halos \citep{Springel08}. The start redshift of the
Aquarius halos was 127, so even for a spherical top-hat collapse the
timing error given by equation~(\ref{zeld_delay}) would only be around
90 Myrs for Zel'dovich initial conditions, which is short compared to
the age of the Universe, so any effects on the main halo are expected
to be small.

  We chose the Aq-A-5 halo, with about 600000 particles within
$r_{200}$, from \cite{Springel08} as a convenient halo to study the
effects of varying the start redshift and initial conditions.  For
this study, we ran eight (re)simulations of Aq-A-5 to redshift zero,
starting at redshifts 127, 63, 31 and 15, for both Zel'dovich and 2lpt
initial conditions.  We used the P-Gadget3 (which is needed to create
the 2lpt initial conditions) and took identical values for the various
numerical parameters as in \cite{Springel08}, and similarly ran the
SUBFIND algorithm to identify subhalos. At this resolution, there are
typically about 230 substructures with more than 31 particles in the
main halo.

\begin{figure}
\resizebox{\hsize}{13.2truecm}{
\includegraphics{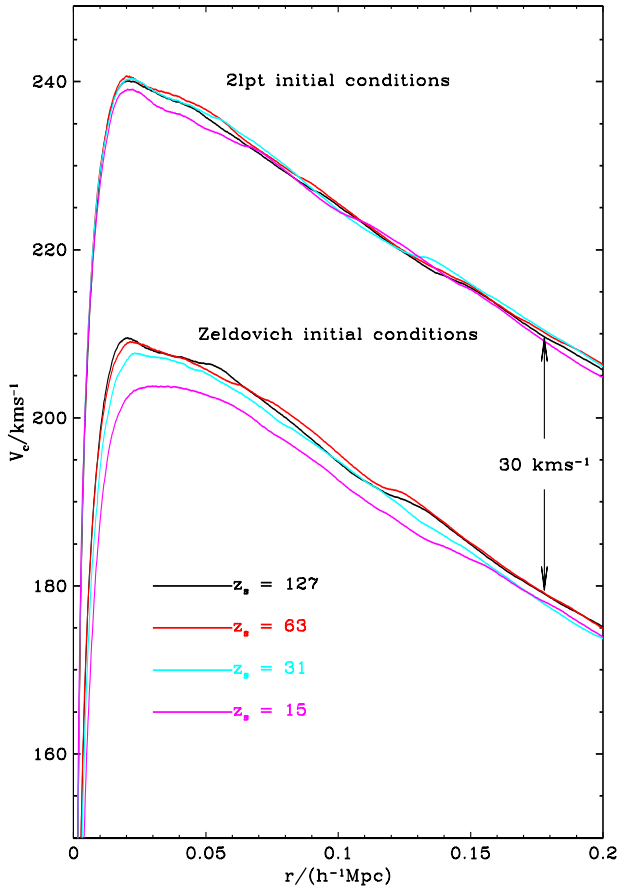}}
\caption{ The circular velocity as a function of radius for the
Aquarius Aq-A-5 halo at z=0 for simulations starting from 2lpt initial
conditions (shifted vertically by 30 kms$^{-1}$ for clarity) and
Zel'dovich initial conditions, starting from the redshifts indicated in
the key.  The curves for 2lpt initial conditions are much more
consistent amongst themselves.  The Aquarius Aq-A-5 halo in
Springel~et~al.~(2008) 
was started at $z_s = 127$ from Zel'dovich initial
conditions. In the absence of the 30 kms$^{-1}$ vertical shift, the
$z_s=127$ curve starting from Zel'dovich initial conditions lies
between or very close to the 2lpt circular velocity curves.}
\label{figvc}
\end{figure}

  In Figure~\ref{figvc} we show the circular velocity as a function
of radius for Aq-A-5 halo at redshift zero for all eight simulations.
The radius is taken from the position of the potential minimum
identified by SUBFIND \citep{Springel01}, and the circular velocity is
computed assuming spherical symmetry for the mass distribution about
the potential minimum.  Comparing the circular velocities for the
Zel'dovich initial initial conditions amongst themselves, and likewise
the 2lpt circular velocities (displaced vertically by 30 kms$^{-1}$ for
clarity) it
is first evident that the circular velocity curves are almost all
rather alike, with the simulation starting from $z_s=15$ Zel'dovich
initial conditions being the outlier. For these initial
conditions the peak of the circular velocity at $z=0$ 
occurs almost 50\% further from the centre than in the other
simulations.  Comparing the curves it is clear that the circular
velocity shows a much weaker dependence on the start redshift with the
2lpt initial conditions than with Zel'dovich initial conditions.

\begin{figure}
\resizebox{\hsize}{!}{
\includegraphics{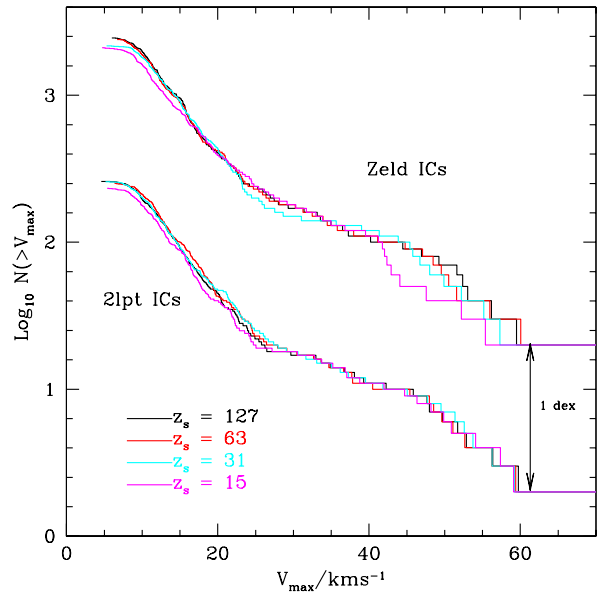}}
\caption{ Cumulative subhalo abundance as a function of the maximum
 subhalo circular velocity for Aquarius halo Aq-A-5 run starting at
 redshift, $z_s$, with Zel'dovich and 2lpt resimulation initial conditions.
 The results from the 2lpt initial conditions are less sensitive to
  the start redshift.  The cumulative function for the Zel'dovich initial
  conditions have been shifted upwards by one dex for clarity.  The
  curve obtained by plotting the Zel'dovich $z_s = 127$ with the shift
  lies within or close to the curves from the 2lpt initial conditions.}
\label{fig_cuml_sub_200}
\end{figure}

  In Figure~\ref{fig_cuml_sub_200} we plot the cumulative subhalo abundance as
a function of the maximum subhalo circular velocity for the eight simulations,
shifting the Zel'dovich curves upward by one dex for clarity. Again, the subhalo
circular velocity functions show a much greater consistency at z=0 amongst the
simulations starting from  2lpt initial conditions compared with Zel'dovich initial 
conditions.

 Finally in this Subsection, we look at the sensitivity of the
position of subhaloes inside the Aq-A-5 halo, as determined by
SUBFIND, to the start redshift and kind of initial conditions.  A
comparison of the relative subhalo positions for the Aq-A halo
resimulated at different resolutions was presented in
\cite{Springel08}. For these simulations it proved possible to
identify corresponding substructures in the different simulations and
compare their relative positions.  All of these simulations were
started from the same start redshift with Zel'dovich initial conditions
so the details of the comparison are not directly relevant,
nonetheless the fact that differences of similar magnitude to those
found here occur, implies that there are other numerical issues which
also affect the positions of substructures significantly.

We can likewise match substructures between our
simulations, and this is made easier by the fact that there is a one
to one correspondence between the dark matter particles in the
different resimulations based on their Lagrangian coordinates.  To
make the a fair comparison between all 28 pairs of simulations we
first found a subset of matched subhalos common to all eight
resimulations.  This sample has only 50 members. This compares to
between 102 and 180 matches between the different pairs of
simulations. The failure to match can arise from two boundary effects:
a subhalo in one simulation may be a free halo in the other; a
`subhalo' may be below the limit of 32 particles in one of the
simulations. Almost half of the subhalos found have less than 64
particles.

\begin{figure}
\resizebox{\hsize}{!}{
\includegraphics{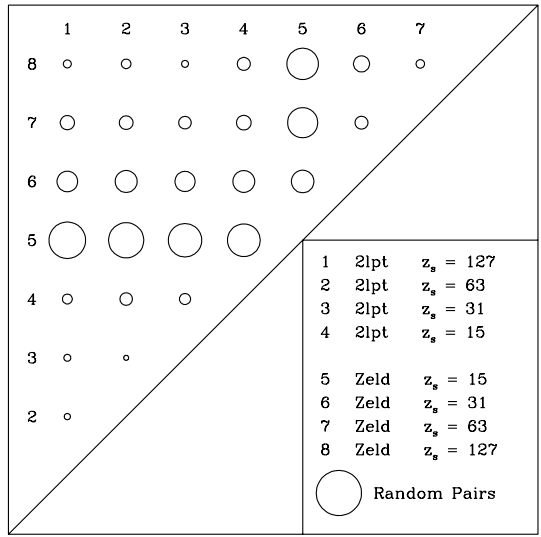}}
\caption{ Sensitivity of the subhalo positions to the start redshift and
type of initial condition.  Each circle above the diagonal represents
the result of a comparison, made at z=0, between two resimulations out
of a total of eight resimulations of halo Aq-A-5. The type of initial
condition and the start redshift, $z_s$, are given in the figure
legend.  The radius of each circle is equal to the geometric mean
distance between a sample of matched subhalos, common to all eight
resimulations. For scale the circle in the legend has a radius of
$192{\rm kpc}/h$ which corresponds to the geometric mean separation,
averaged over all eight resimulations, between randomly chosen pairs
of subhalos within the same subhalo.  The 2lpt initial conditions are
significantly less sensitive to the start redshift than Zel'dovich
initial conditions.  }
\label{subhalo_pos}
\end{figure}  

 The subhalo positions in two different simulations are compared,
after first subtracting the position of the main subhalo.  There are
many ways to compare the differences found.  We have selected to show
the geometric mean of the distance between matched subhalos between
pairs of simulations in Figure~\ref{subhalo_pos}. Comparisons based on
other quantities such as the mean or median separation give
qualitatively similar results.
 
 Again we find, from Figure~\ref{subhalo_pos}, that the simulations
started from 2lpt initial conditions show much less variation in
position on the start redshift than Zel'dovich initial conditions.  For
the simulation with Zel'dovich initial conditions starting from $z_s =
15$, the location of the subhalos is very poorly correlated with any
other of the simulations, although still better than the geometric
mean separation of all non-identical pairs of subhalos in any given
halo.  Starting the Zel'dovich initial conditions at $z_s = 127$, the
starting redshift for the Aquarius halos, does give subhalo positions
which match the 2lpt resimulation subhalo positions fairly well.  We
checked that the analogous plot produced using the larger matched subhalos
between each pair, rather than the subsample of subhalos common to all
eight simulations, gives a very similar plot to the plot shown.

 In conclusion, for a resimulation of a Milky-Way like dark matter
halo it is clear that the $z=0$ properties of the halo do depend
somewhat on the start redshift, but that dependence is much reduced
with 2lpt resimulation initial conditions.  The evolution of the halo
to $z=0$ is highly non-linear, but the results are qualitatively in
agreement with the known behaviour of 2lpt and Zel'dovich initial
conditions in the quasi-linear regime \citep{Scoccimarro98} where this
can be understood from the behaviour of transients which are both
smaller and decay faster (by a factor $1+z$) for 2lpt initial
conditions than for Zel'dovich initial conditions. In quantitative terms
the differences amongst the circular velocity curves, the cumulative
subhalo abundances and the relative positions of substructures between
the different simulations started from 2lpt initial conditions, is
larger than na\"ively expect from the behaviour of transients.  This
is most likely because there are other numerical factors which lead to
much of the differences seen in the various simulations starting from
2lpt initial conditions.

  To show this is indeed the case we repeated one resimulation using
initial conditions that were mirror reflected about the centre of the
periodic volume. Running these otherwise identical simulations leads,
after reflecting the final conditions, to differences in the positions
of the subhalos. The numerical process of mirroring is not a perfect
symmetry operation when operated on double precision floating point
numbers so in effect the comparison shows that the effects of rounding
error in the positions can propagate to much larger scales.  After
matching 189 subhalos for the z=0 halos we find the geometric
separation of the subhalos between these reflected initial conditions
is 12 kpc$/h$. This can be compared to 26 kpc$/h$ difference between
the 175 matched subhalos between the 2lpt resimulations starting at
redshifts 127 and 63.  This kind of sensitivity to the initial
conditions is expected given that halo formation is a strongly
non-linear processes.  It is interesting however, even though these
simulations of halo formation show this kind of sensitivity to small
changes in the initial conditions, that the effects of starting from
2lpt or Zel'dovich initial conditions are still apparent at redshift
zero for many properties including for the circular velocity curve,
the subhalo velocity functions and subhalo positions.

\section{Summary and Discussion\label{SECTSUMMARY}}

  We have implemented a new method to make second order Lagrangian
perturbation theory (2lpt) initial conditions which
is well suited to creating resimulation initial conditions. We have
tested the method for an analytic spherically symmetric perturbation
with features ranging over a factor of 5000 in linear scale or
eleven orders of magnitude in mass, and can reproduce the analytic
solution to a fractional accuracy of better than one percent over
this range of scales in radius measured from the symmetry centre.

 Applying the new method we have recreated the initial conditions for
the formation of a high redshift halo from \cite{Gao05} and the a
Milky-way mass halo from \cite{Springel08}.  We find from studying the
properties of the final halos, that the final conditions show much
less sensitivity to the start redshift when using 2lpt initial
conditions than when started from Zel'dovich initial conditions.

 For didactic purposes, we have calculated the effect of using
Zel'dovich and 2lpt initial conditions for a spherical top-hat
collapse. In this simple case, the epoch of collapse is delayed by a
timing error which depends primarily on the number of expansions
between the epoch of the initial conditions and the collapse
time. This timing error, for fixed starting redshift, grows with
collapse time for Zel'dovich initial conditions, but decreases with
collapse time for 2lpt initial conditions. Applying the
top-hat timing errors as a correction to the radius of shells 
containing fixed amounts of mass as a function of time has the
effect of bringing the Zel'dovich and 2lpt resimulations of the 
high redshift \cite{Gao05} halo into near coincidence.

  We find that for the Aq-A-5 halo the choice of start redshift for
the Zel'dovich initial conditions in \cite{Springel08} was
sufficiently high at $z_s=127$ to make little difference to the
properties of the halo at $z=0$ when compared to runs starting from
2lpt initial conditions.  For lower starting redshifts ($z_s = 63, 31,
15$) the positions of substructures at $z=0$ do become sensitive to
the precise start redshift.

  \cite{Knebe09} have also looked at the properties of halos, for the
mass range ($10^{10} - 10^{13}h^{-1}M_\odot$) at redshift zero, for
simulations starting from Zel'dovich and 2lpt initial conditions
(generated using the code by \cite{Crocce_etal06}).  In their paper
they simulated a cosmological volume which yielded many small halos
and about ten halos with 150000 or more particles within the virial
radius. In their study they looked at the distribution of halo
properties including the spin, triaxiality and concentration.  They
also matched halos in a similar way to the way we have matched
subhalos and looked at the ratios of masses, triaxialities, spins and
concentrations of the matching halos. The authors concluded that any
actual trends with start redshift or type of initial condition by
redshift zero are certainly small (for start redshifts of 25 - 150).
It is not possible to directly compare our results with theirs, but
the trends in the halo properties we have observed in halo Aq-A-5 at
redshift zero are indeed small, and not obviously in disagreement with
their results for populations of halos. 

  For many purposes, the differences which arise from using Zel'dovich
or 2lpt initial conditions may be small when compared to the
uncertainties which arise in the modelling of more complex physical
processes.  From this point of view the main advantage of using 2lpt
initial conditions is that they allow the simulations to start later
thus saving computer time. The issue of exactly what redshift one can
start depends on the scientific problem, but the spherical collapse
model, discussed in Subsection~\ref{substh} gives a quantitative way
to compare the start redshift for Zel'dovich or 2lpt initial
conditions.  However it remains the case that it is advisable to test
the sensitivity of simulation results to the start redshift by direct
simulation as for example in \cite{Power03} for resimulations of
galactic dark matter halos.

\section*{ACKNOWLEDGEMENTS} I would particularly like to thank Volker
Springel for implementing part of the method described in this paper
in his P-Gadget3 code and for reading a draft of this paper.  I would
also like to thank Roman Scoccimarro for clarifying an issue connected
with the second-order overdensity field early on in this project, Gao
Liang, for helping with remaking the \cite{Gao05} initial conditions
and Carlos Frenk for comments on the first draft. The simulations in
this paper were run on COSMA and I am grateful to Lydia Heck for
computer support. This work was supported by STFC rolling grant
ST/F002289/1.

\bibliography{2lpt}

\section{Appendix: Methods for interpolation of displacement fields}

As mentioned in the main text, it is necessary to use interpolation of
the displacement fields when making resimulation initial conditions in
order to find the displacements at the particle positions. This need
arises for example because the interparticle spacing in the high resolution
region of the resimulation initial conditions is typically many times
smaller than the Fourier grid used to generate the longest wavelength
displacements over the entire periodic simulation volume. Another
situation where interpolation is needed, even for cosmological initial
conditions, occurs when the particle load is based on a glass-like
distribution \citep{White96} which means the particles cannot coincide
with the grid points.

 In this Appendix we will examine two interpolation methods which are
in use and compare them.  We will simplify the discussion to the case
in the limit of the interparticle separation going to zero creating a
continuous and uniform density medium. We assume that the displacement
field is built of a finite number of plane waves and that the
displacements are known precisely on a regular cubic grid of points.
The purpose of the interpolation is to reconstruct the displacement
field everywhere between these grid points. The success of a
particular interpolation scheme can be judged by comparing the density
field created by applying the displacements given by interpolated
displacement field with the density field generated from the true
displacement field. We will assume the displacements are small so
there is no shell crossing and further consider only the linear
dependence of the density field on the displacements.  The interpolation
schemes we consider use linear combinations of the values known at the
grid points, so it is sufficient to characterise the interpolation
scheme by applying it to plane waves.

 We will compare the performance of two interpolation schemes:  the
first, trilinear interpolation or Cloud-in-Cell or CIC,
\cite{Hockney88}, has been used for making initial conditions over
many years for example in the Hubble Volume
simulations\citep{Evrard02} and the Millennium simulation
\citep{Springel05} as well as resimulation initial conditions e.g.
\cite{Gao05}. The second method, not previously documented, which we
call QI, for quadratic interpolation, has been used more recently for
the Aquarius initial conditions \citep{Springel08}.

  In principle, because the displacement fields for resimulations are
generated by Fourier methods are bandwidth limited, the interpolation
of the displacement field to the particle positions can be done
perfectly using the Whittaker-Shannon interpolation formula
\citep{Numerical_recipes92}.  This is impractical because the sum must be taken
over all grid points.  The only exception to this is the trivial case
where the particles lie on the Fourier grid points themselves, in
which case no interpolation is needed. For cosmological initial
conditions, unlike resimulation initial conditions, this can easily be
arranged, as for example in the 2lpt initial condition code made
available by \cite{Crocce_etal06}.

 Before discussing particular interpolation formulae we first define
some convenient notation.  Suppose the computational domain is a
periodic cube of side length $L$ and that we define a cubic grid with
$p^3$ cubic cells of side length $\Delta$ such that $L=p\Delta$.  A
general point in the volume has coordinates $(x,y,z)$ where all 
three spatial coordinates have a range $0 \le x,y,z <L$. We will
consider the interpolation of a scalar field $A(x,y,z)$ and we will
assume $A$ a bandwidth limited function, in such a way that its value
can be predicted exactly anywhere within the domain using the
Whittaker-Shannon interpolation formula and summing over the $p^3$ grid points.

We first consider trilinear interpolation. The interpolated field,
$A_{\rm TLI}$, is given by the expression:

\begin{equation}\label{app-eq1}
A_{\rm TLI}= \sum_{l_1,l_2,l_3} W(\frac{x}{\Delta}-l_1) W(\frac{y}{\Delta}-l_2) W(\frac{z}{\Delta}-l_3)A,
\end{equation}
where the $(l_1,l_2,l_3)$
represents an integer triplet and the sum is over all $p^3$ grid
points. For TLI the kernel $W$ is defined as:
\begin{equation}
W(a) = 
\begin{cases}  1 - |a| &\text{if $|a| \le 1$} \\
               0       &\text{if $|a|  >  1$}.
\end{cases}
\end{equation}

  TLI interpolation is exact for a function which is a sum of terms
 $x^{\alpha_1}y^{\alpha_2}z^{\alpha_3}$ where $0\le\alpha_i\le1$.  The
 compact form of $W$ means that to evaluate the interpolated field at
 the position of a particle requires summing only eight terms
 corresponding to the nearest grid points.  The resulting field,
 $A_{\rm TLI}$, is continuous everywhere, which is a necessary
 requirement when interpolating a component of the displacement field,
 otherwise regions of zero or double density can be created along the
 interfaces marking the discontinuities. When TLI is applied to the
 components of the displacement field, the resulting density field has
 discontinuities.

  We will now consider the effect on a plane wave. The `interpolated'
wave will no longer be a simple plane wave but a superposition of the
`fundamental' plane wave (with a reduced amplitude) and the harmonics of the
fundamental wave.  The same is true of the density field generated by
applying the interpolated displacements (assumed normal to the plane
wave and of small amplitude) to a uniform medium.  The relative power
in the fundamental wave and the harmonics will differ between  the
displacement field and the density field, with relatively more power in the density
field harmonics. The ideal interpolation scheme minimises the power in
these harmonics and maximises the power in the fundamental plane wave.

 Consider a plane wave mode in a periodic cube of side-length
$L$ :
\begin{equation} 
 A(k_1,k_2,k_3) = B \exp[i\frac{2\pi}{L}(k_1x+k_2y+k_3z)],
\end{equation}
where $k_1$, $k_2$, $k_3$ are integers and $B$ is the amplitude of
the wave.
 It is straightforward to show that applying TLI interpolation to
the plane wave gives a reduced amplitude, $B_{\rm TLI}$, for 
the fundamental plain wave given by the following ratio:
\begin{equation}
          B_{\rm TLI}/B =
 \left(\frac{\sin u_1}{u_1}\right)^2
 \left(\frac{\sin u_2}{u_2}\right)^2
 \left(\frac{\sin u_3}{u_3}\right)^2,
\label{tli_red}
\end{equation} where $u_i = \pi k_i/p$ provided $0<u_i\le\pi/2$.

  This factor applies equally to the displacement field and
the linear density field created by
applying the interpolated displacement field to a uniform medium.  For
small values of $u_i$ the ratio is close unity and deviates below
unity quadratically to lowest order.  It is desirable to use as large
a Fourier grid as memory allows for making resimulation initial
conditions to minimise the interpolation error.

  The interpolation can be made more accurate if the derivatives of
the field, $A$, are also known at the grid points.  The derivatives of
the displacement field need to be computed for 2lpt initial conditions
in any case and it was the fact that the derivatives need to
be calculated anyway that motivated trying to find an improved interpolation
scheme making use of them.

  It is straightforward, making use of the derivatives of the field
to devise a formula which is exact for linear and quadratic functions
of the coordinates.  We will call it the quadratic scheme. The field
$A_{\rm QI}$ is given by:
\begin{equation}
\begin{aligned}
  A_{\rm QI} = & A_{\rm TLI} + \\
  &  \sum_{l_1,l_2,l_3} V(\frac{x}{\Delta}-l_1)W(\frac{y}{\Delta}-l_2) W(\frac{z}{\Delta}-l_3)\frac{\partial A}{\partial x} + \\
  &  \sum_{l_1,l_2,l_3} W(\frac{x}{\Delta}-l_1)V(\frac{y}{\Delta}-l_2) W(\frac{z}{\Delta}-l_3)\frac{\partial A}{\partial y} + \\
  &  \sum_{l_1,l_2,l_3} W(\frac{x}{\Delta}-l_1)W(\frac{y}{\Delta}-l_2) V(\frac{z}{\Delta}-l_3)\frac{\partial A}{\partial z}, 
\end{aligned}
\end{equation}
where:
\begin{equation}
V(a) = 
\begin{cases}  \phantom{-}a(1-a)/2 &\text{if $0 \le a \le 1$} \\
                         -a(1+a)/2 &\text{if $-1 \le a \le 0$} \\ 
               \phantom{-}0        &\text{if $|a|  >  1$}.
\end{cases}
\end{equation}
 The QI interpolation is exact  for any function which is a sum of
terms $x^{\alpha_1}y^{\alpha_2}z^{\alpha_3}$ where $0\le\alpha_i\le2$.
The three additional terms in the QI formula vanish at the grid
points so like TLI interpolation formula, the QI scheme is exact
at the grid points.  The corresponding effect of QI interpolation on a plane
wave can be shown to be:
\begin{equation}
\begin{aligned}
 &          B_{\rm QI}  /   B      = 
 \left(\frac{\sin u_1}{u_1}\right)^2
 \left(\frac{\sin u_2}{u_2}\right)^2
 \left(\frac{\sin u_3}{u_3}\right)^2  + \\
 & \left[\left(\frac{\sin u_1}{u_1}\right)^2-\frac{\sin 2u_1}{2u_1}\right]
\left(\frac{\sin u_2}{u_2}\right)^2 \left(\frac{\sin u_3}{u_3}\right)^2 \\
 & \left[\left(\frac{\sin u_2}{u_2}\right)^2-\frac{\sin 2u_2}{2u_2}\right]
\left(\frac{\sin u_3}{u_3}\right)^2 \left(\frac{\sin u_1}{u_1}\right)^2 \\
 & \left[\left(\frac{\sin u_3}{u_3}\right)^2-\frac{\sin 2u_3}{2u_3}\right]
\left(\frac{\sin u_1}{u_1}\right)^2 \left(\frac{\sin u_2}{u_2}\right)^2 \\
\label{qli_red}
 \end{aligned}
\end{equation} 
where $u_i = \pi k_i/p$ provided $0<u_i\le\pi/2$.  Again this ratio is
  close to unity, when $u_i$ is small, but the leading order deviation
  below unity for small $u_i$ is now quartic - a considerable improvement.

\begin{figure}
\resizebox{\hsize}{!}{
\includegraphics{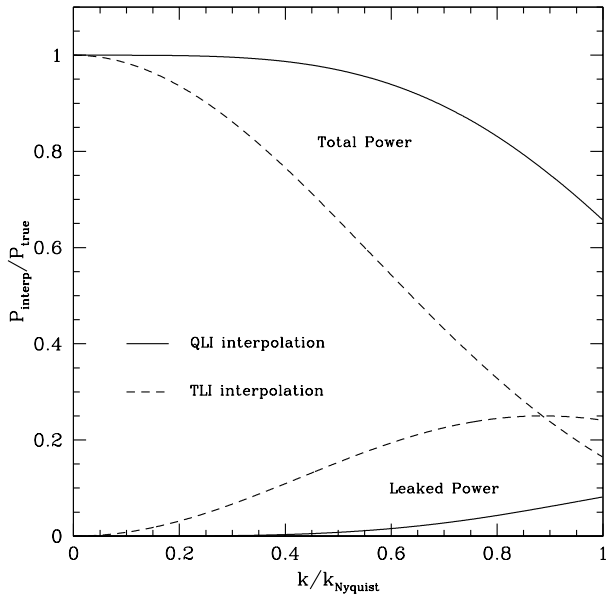}}
\caption{ The curves show the effects of the TLI and QI interpolation
schemes when applied to a 1-d on a plane wave, representing a
displacement field normal to the wave, of wavenumber $k$, given a
sampling rate of the plane wave of wavenumber $k_{\rm Nyquist}$.  The
upper curves show the power remaining in resulting linear density
created by displacing a uniform medium, with the plane wave
displacements given by interpolation, while the lower curves show the
power introduced into the the harmonics of the plane wave by the
process of interpolation. }
\label{interpfig}
\end{figure}

  Figure~\ref{interpfig} compares the two interpolation schemes acting
on a plane wave along one of the principal coordinate axes.  The upper
curves show the ratio of the power in the density field created by applying
the interpolated displacements with the true power for both TLI and QI
interpolation schemes. The curves are obtained by setting $u_2 = 0;
u_3 = 0$ in equations~(\ref{tli_red}) and (\ref{qli_red}) respectively
and the x-axis is plotted up to the Nyquist wavenumber of the Fourier
grid points where the displacement field is known exactly.  Both
interpolation schemes introduce power into the harmonics of the
plane wave.  For a 1-d plane wave the TLI scheme this harmonic power
can be shown to be:
\begin{equation}
 \frac{P_{\rm harm}}{P_{\rm true}} = \left(\frac{\sin u}{u}\right)^2-\left(\frac{\sin u}{u}\right)^4,
\end{equation}
while for the QI scheme the harmonic power is given by:
\begin{equation}
\frac{P_{\rm harm}}{P_{\rm true}} =  \left(\frac{\sin u}{u}\right)^2\left(1+\frac{u^2}{3}\right)-
             \left(\frac{2\sin^2 u}{u^2}-\frac{\sin 2u}{2u}    \right)^2,
\end{equation}
where $u = k\pi/p$ provided $0<u\le\pi/2$.  The lower two curves show
 the amount of power generated in the harmonics for the TLI and QI
 interpolation schemes.  The QI scheme is evidently superior to TLI on
 both measures.

 Both the TLI and QI interpolation schemes have cubic symmetry, and
therefore generate some anisotropies. However in both schemes these
anisotropies are small and enter first at quartic order in the
magnitude of the wavevector. The fractional RMS derivation in 
the power in the fundamental wave about the average over
all directions less than 1 percent even at the Nyquist wavelength for both TLI and QI,
with QI being the best by a factor three.

 For a fixed grid size implementing the QI formula is a factor of
four more expensive than TLI as the three derivatives of $A$ need to
be computed in addition to $A$ itself. For initial conditions, this
factor is reduced to three, because the quantities of interest are
gradients of a potential which reduces the number of independent terms
that have to be evaluated.  The QI scheme however is more attractive
after taking account of its improved accuracy and the fact that the computational effort 
scales approximately as the cube of the dimension of the Fourier grid.
For limited memory or limited CPU the QI scheme is to be
preferred to TLI.

 The QI interpolation scheme was used for making all the Zel'dovich
and 2lpt initial conditions used in this paper.  For the derivatives
of the Zel'dovich displacements, which are required to compute the
second-order overdensity, using the QI interpolation scheme entails
calculating the second order derivatives of the Zel'dovich
displacements.

\end{document}